\documentclass{iaus}
\usepackage{graphicx}

\title[Galactic C and S Stars as Guidelines for Magellanic Cloud AGB Stars] 
{Galactic C and S Stars as Guidelines for Magellanic Cloud AGB
Stars}

\author[Roald Guandalini \& Maurizio Busso]   
{Roald Guandalini$^1$ \and Maurizio Busso$^1$}

\affiliation{$^1$Dipartimento di Fisica, Universit\`{a} di Perugia
and INFN, Sezione di Perugia,
\\ via A. Pascoli, 06123, Perugia, Italy \\ email:
{\tt guandalini@fisica.unipg.it, busso@fisica.unipg.it}}

\pubyear{2008}
\volume{256}  
\pagerange{119--126}
\jname{The Magellanic System: Stars, Gas, and Galaxies}
\editors{Jacco Th. van Loon \& Joana M. Oliveira, eds.}
\begin{document}

\maketitle

\begin{abstract}
The study of the evolutionary properties of Asymptotic Giant
Branch stars still presents unresolved topics. Progress in the
theoretical understanding of their evolution is hampered by the
difficulty to empirically explain key physical parameters like
their luminosity, mass loss rate and chemical abundances. We are
performing an analysis of Galactic AGB stars trying to find
constraints for these parameters. Our aim is of extending this
analysis to the AGB stars of the Magellanic Clouds and of the
Dwarf Spheroidal Galaxies using also mid-infrared observations
from the Antarctic telescope IRAIT. AGB sources from the
Magellanic Clouds will be fundamental in our understanding of the
AGB evolution because they are all at a well defined distance
(differently from the Galactic AGBs). Moreover, these sources
present different values of metallicity: this fact should permit
us of examining in a better way their evolutionary properties
comparing their behaviour with the one from Galactic sources.
\keywords{stars: AGB and post-AGB, stars: evolution, infrared:
stars}
\end{abstract}

\firstsection 
\section{Introduction}

The Asymptotic Giant Branch (hereafter AGB) phase lies at the end
of the active life for Low- and Intermediate-Mass Stars (M $<$ 8
Solar Masses); a detailed description can be found in \cite[Busso
et al. (1999)]{Busso_etal99} and references herein. Stars in this
phase lose mass very effectively; stellar winds deeply affect
their evolution, and are fundamental for the C-enrichment of the
Interstellar Medium. Moreover, AGB winds account for about 70$\%$
of all the matter returned after stellar evolution (\cite[Sedlmayr
1994]{Sedlmayr94}) and provide the starting conditions for the
formation of planetary nebulae. Radiation from cool dust particles
in the infrared (IR) normally dominates the energy distribution of
AGB stars: this fact is due to their strong stellar winds building
up huge dusty circumstellar envelopes.

Our quantitative knowledge of crucial chemical and physical
parameters of AGB sources is unfortunately still poor. Among the
uncertain issues we emphasize in particular the mass loss rate and
the distance, hence also the luminosity. These facts have hampered
for decades our capability of satisfactorily describing the
physics of these dust-enshrouded variable objects, despite their
evolution is based on the two best known nucleosynthesis phases,
namely H and He burning.

\section{A Study of Galactic AGBs}

\begin{figure}[t]
\begin{center}
\includegraphics[width=4in]{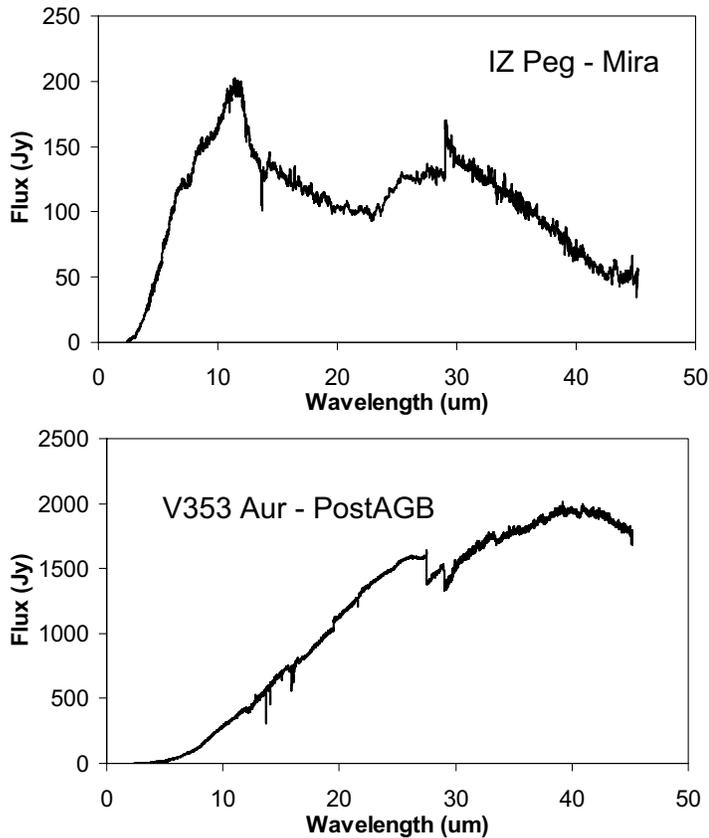}
\caption{ISO-SWS1 spectra of 2 C-rich evolved AGB stars
(\cite[Guandalini et al. 2006]{Guandalini_etal06}).}
   \label{fig1}
\end{center}
\end{figure}

\begin{figure}[t]
\begin{center}
{\includegraphics[width=2.3in,
height=2.3in]{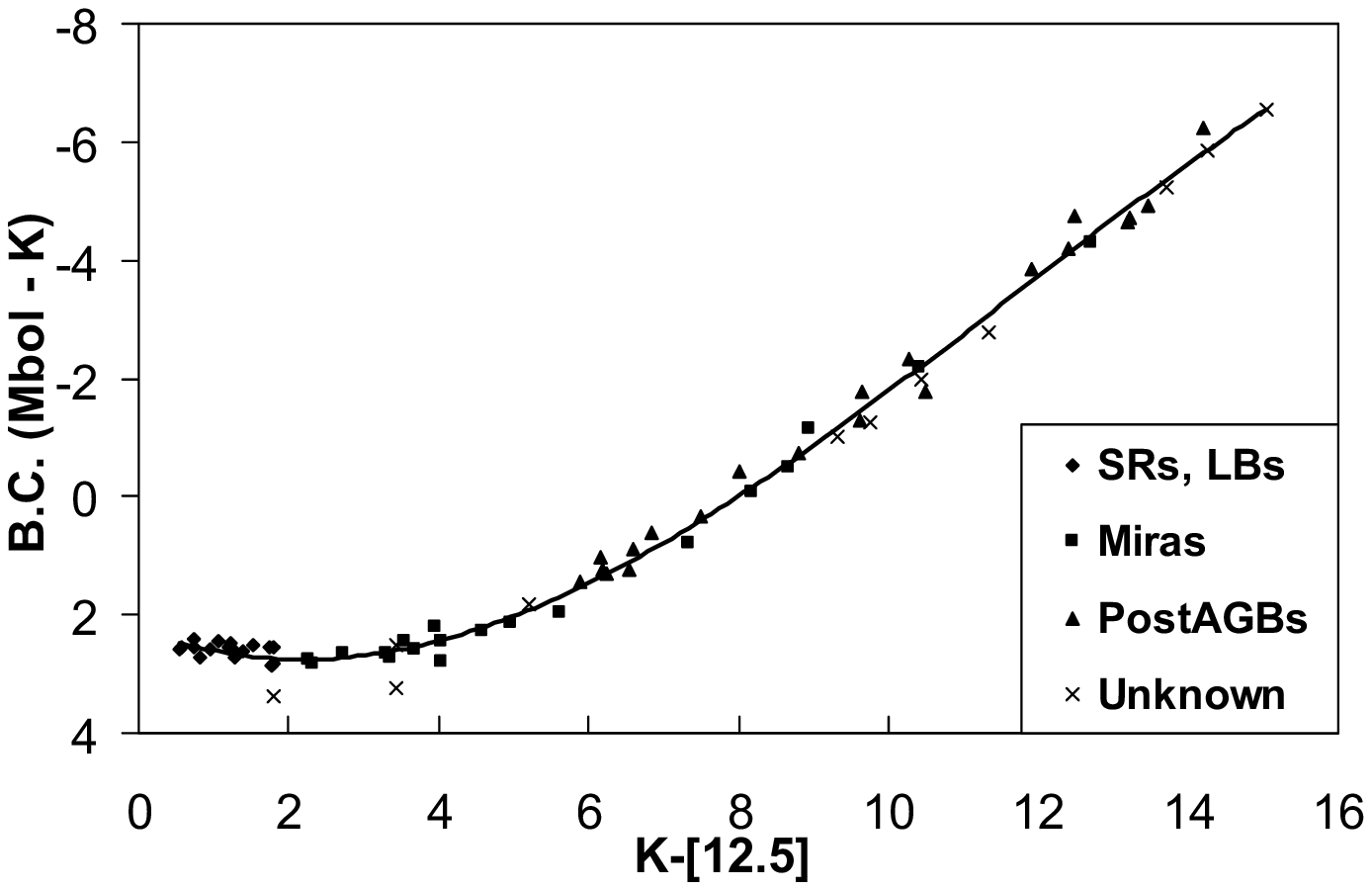}\includegraphics[width=2.3in,
height=2.3in]{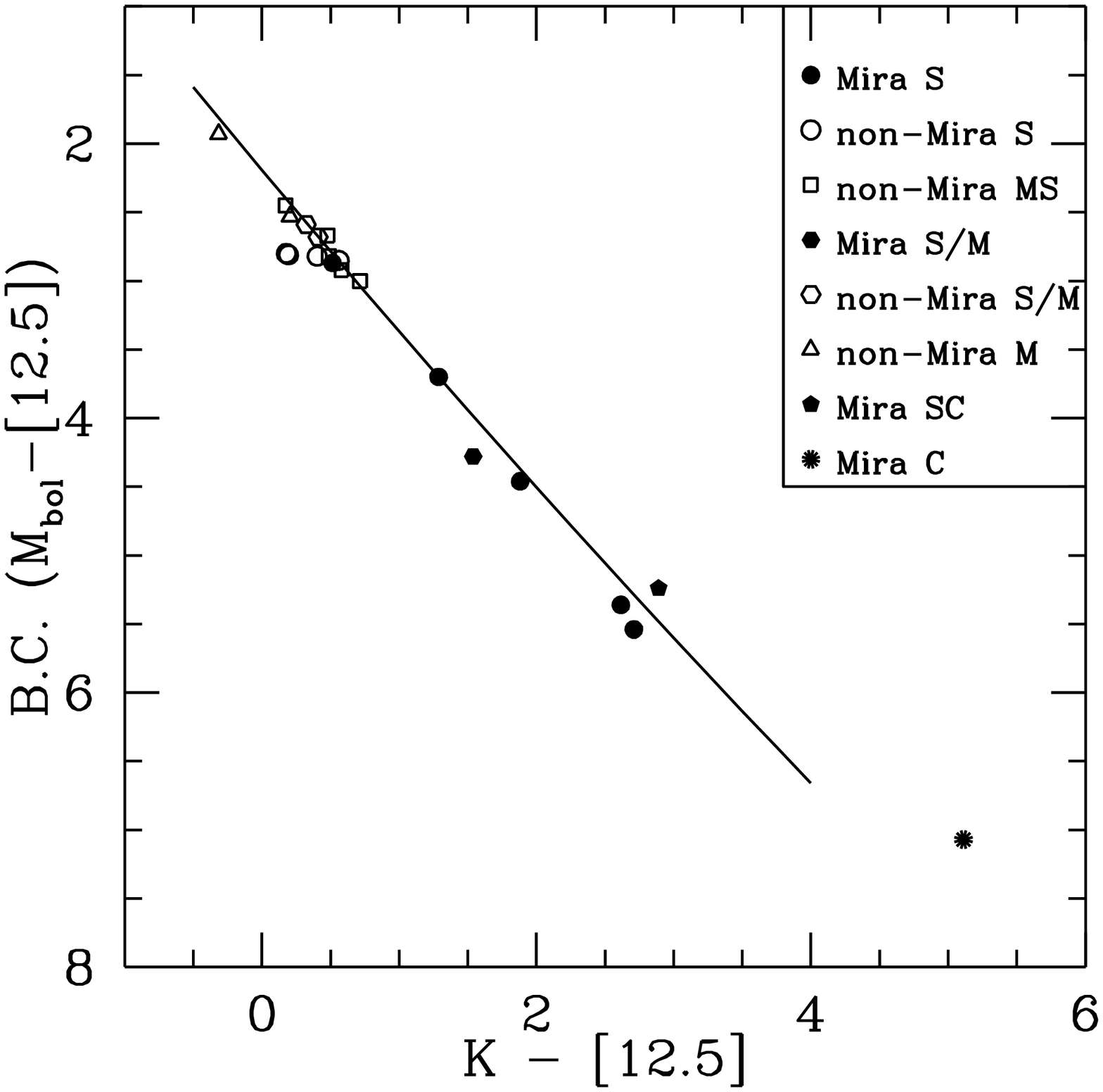}}
\caption{Bolometric Corrections for samples of C-rich (left) and
S-type (right) AGB stars.}
   \label{fig2}
\end{center}
\end{figure}

We are performing an analysis of Galactic AGB stars looking for
correlations between their basic parameters (bolometric
luminosity, mass loss rate, etc... ) and observable quantities.
Our main aim is that of establishing observationally-based
criteria permitting a more quantitative determination of mass loss
rates and of luminosities for the various types of AGB stars, thus
providing general rules to be adopted in the improvement of
stellar codes. Extensive samples of well studied Galactic AGB
stars (M-type, S-type, C-rich) have been collected. They are large
enough (several hundreds of sources per type) that conclusions on
them have a good statistical significance. They also have detailed
and accurate Spectral Energy Distributions (SEDs) at near- and
mid-IR wavelengths and, very often, reliable measurements of mass
loss rates and distances. The first results of this ongoing
research have been already published for carbon-rich and S-type
stars (\cite[Guandalini et al. 2006]{Guandalini_etal06},
\cite[Busso et al. 2007]{Busso_etal07}, \cite[Guandalini \& Busso
2008]{GuaBus08}).

\section{The Importance of Infrared Observations}

\begin{figure}[t]
\begin{center}
 \includegraphics[width=3.7in, angle=-90]{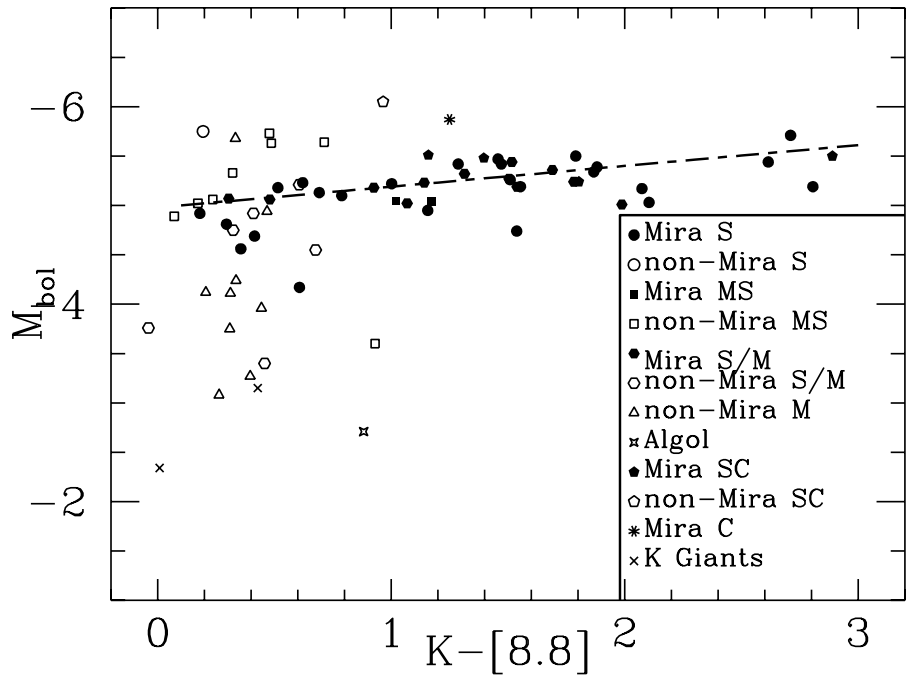}
 \caption{HR diagram of a sample of S-type stars (\cite[Guandalini
\& Busso 2008]{GuaBus08}).}
   \label{fig3}
\end{center}
\end{figure}

An extended wavelength coverage is fundamental in the study of the
evolutionary phases of AGB stars. In order to show this,
Fig.\,\ref{fig1} displays the SEDs of evolved AGBs (usually
Miras), which emit a large part of their flux at mid-IR
wavelengths. As a consequence, both near and mid-IR observations
sources are necessary to obtain good estimates of the apparent
bolometric magnitudes, either by physically integrating the SEDs,
or by applying pre-calibrated, reliable bolometric corrections
(B.C.). Examples of such corrections are presented in
Fig.\,\ref{fig2} as a function of infrared colours (see
\cite[Guandalini et al. 2006]{Guandalini_etal06}, \cite[Guandalini
\& Busso 2008]{GuaBus08}). Once mid-IR wavelengths and bolometric
corrections have been properly included, an example of the
(absolute) HR diagrams that can be obtained is shown in
Fig.\,\ref{fig3}. The straight dashed line illustrates how, at
least for Mira variables, a rather well defined relation emerges
between the absolute bolometric magnitude and a near-to-mid
infrared color (this last also directly linked to the extension
and temperature of the circumstellar envelope).

\section{AGB Stars and Magellanic Clouds}

The study of the AGB phase through the analysis of Galactic
sources presents many problems. As an example we can remember the
large difficulties in estimating their distances. AGB sources from
the Magellanic Clouds will therefore be fundamental in our
understanding of the AGB evolution, because for them the problem
of obtaining reliable distances is avoided by the good knowledge
of the distance modulus. The above fact allows a good estimate for
the Luminosity in each photometric band, so that our bolometric
corrections would provide the bolometric magnitudes directly.
Moreover, AGB sources of the two Magellanic Clouds have different
values of the metallicity, both with respect to each other and
with respect to the Galaxy. This will allow us to study global
properties (Luminosity, mass loss rate, ratio of the number of C
stars to M giants) also as a function of the chemical composition.
Our aim is therefore to extend the analysis, which has been almost
completed for Galactic AGB stars, to the Magellanic Clouds and to
close-by Dwarf Spheroidal Galaxies. For doing this, important
tools will be offered by the exploitation of Antarctic Infrared
Astronomy, as offered by the Italo- Spanish robotic telescope
IRAIT (\cite[Tosti et al. 2006]{Tosti_etal06}).

\end{document}